\documentclass[aps,preprint,tightenlines,superscriptaddress,showpacs,byrevtex]{revtex4}

\usepackage{epsfig}  
\usepackage{dcolumn} 
\graphicspath{ps}

\newcommand{\cc}[1]{\multicolumn{1}{c}{#1}}

\newcommand{\hdig}{\hphantom{0}}

\newcommand{\hsa}{\hspace{+1.0mm}}
\newcommand{\hsb}{\hspace{+0.5mm}}

\newcommand{\hse}{\hspace{+0.5em}}
\newcommand{\hsf}{\hspace{+0.2em}}
\newcommand{\hsg}{\hspace{+0.7em}}
\newcommand{\hsh}{\hspace{+0.1em}}

\newcommand{\aer}[2]{\mbox{$^{\hspace{+0.3em}+\hspace{+0.55em}#1}_{\hspace{+0.3em}-\hspace{+0.55em}#2}$}}
\newcommand{\aerb}[2]{\mbox{$^{\hspace{+0.3em}+\hspace{+0.45em}#1}_{\hspace{+0.3em}-\hspace{+0.45em}#2}$}}
\newcommand{\aeraer}[4]{\mbox{$^{\hspace{+0.3em}+\hspace{+0.55em}#1\hspace{+0.4em}+\hspace{+0.55em}#3}_{\hspace{+0.3em}-\hspace{+0.55em}#2\hspace{+0.4em}-\hspace{+0.55em}#4}$}}
\newcommand{\ul}[1]{\mbox{$\hspace{-0.5em}<#1$}}

\def\NsA{\mbox{$595.9\aeraer{33.2}{32.5}{\hdig 7.8}{\hdig 7.7}$}}
\def\SigA{\mbox{$24.1$}}

\def\BrA{\mbox{$18.5\pm 1.0\pm 0.7$}}
\def\NsB{\mbox{$198.9\pm 21.5\aer{15.6}{\hdig 4.8}$}}
\def\SigB{\mbox{$10.8$}}

\def\BrB{\mbox{$12.0\pm 1.3\aerb{1.3}{0.9}$}}
\def\NsC{\mbox{$187.0\pm 16.3\aer{\hdig 1.5}{\hdig 1.7}$}}
\def\SigC{\mbox{$16.4$}}

\def\BrC{\mbox{$22.0\pm 1.9\pm 1.1$}}
\def\NsD{\mbox{$\hdig 72.6\pm 14.0\aer{\hdig 4.9}{\hdig 5.5}$}}
\def\SigD{\mbox{$\hdig 5.8$}}

\def\BrD{\mbox{$11.7\pm 2.3\aerb{1.2}{1.3}$}}

\def\NsE{\mbox{$132.7\aeraer{18.9}{18.2}{\hdig 2.7}{\hdig 2.9}$}}
\def\SigE{\mbox{$\hdig 8.5$}}

\def\BrE{\mbox{$\hdig 4.4\pm 0.6\pm 0.3$}}
\def\NsF{\mbox{$\hdig 72.4\pm 17.4\aer{\hdig 3.7}{\hdig 3.4}$}}
\def\SigF{\mbox{$\hdig 4.5$}}

\def\BrF{\mbox{$\hdig 5.0\pm 1.2\pm 0.5$}}
\def\NsG{\mbox{$\hdig 25.6\aeraer{\hdig 9.3}{\hdig 8.4}{\hdig 1.6}{\hdig 1.4}$}}
\def\SigG{\mbox{$\hdig 3.5$}}

\def\BrG{\mbox{$\hdig 1.7\pm 0.6\pm 0.2$}}
\def\NsH{\mbox{$\hsf -1.0\aer{\hdig 6.6}{\hdig 5.9}$}}
\def\SigH{\mbox{$\hdig 0.0$}}

\def\BrH{\mbox{$\ul{0.7}$}}
\def\NsI{\mbox{$\hdig\hdig  8.6\pm\hdig 5.9$}}
\def\SigI{\mbox{$\hdig 1.6$}}

\def\BrI{\mbox{$\ul{3.3}$}}
\def\NsJ{\mbox{$\hdig\hdig 2.0\pm\hdig 1.9$}}
\def\SigJ{\mbox{$\hdig 1.3$}}

\def\BrJ{\mbox{$\ul{1.5}$}}

\def\nbb{\mbox{$85.0$}}
\def\Nbb{\mbox{$85.0 \pm 0.5$}}

\def\fb{\mbox{fb$^{-1}$}}
\def\bb{\mbox{$B\overline{B}$}}
\def\qq{\mbox{$q\overline{q}$}}
\def\mbc{\mbox{$M_{\rm bc}$}}
\def\de{\mbox{$\Delta E$}}

\def\br{\mbox{${\cal B}$}}

\def\bz{\mbox{$B^0$}}

\def\bp{\mbox{$B^+$}}

\def\ks{\mbox{$K^0_S$}}

\def\dz{\mbox{$D^0$}}
\def\dzb{\mbox{$\overline{D}{}^0$}}
\def\dplus{\mbox{$D^+$}}

\def\kk{\mbox{$K\overline{K}$}}
\def\kppim{\mbox{$K^+ \pi^-$}}
\def\kmpip{\mbox{$K^- \pi^+$}}

\def\kppiz{\mbox{$K^+ \pi^0$}}

\def\kzpip{\mbox{$K^0 \pi^+$}}

\def\kspip{\mbox{$\ks \pi^+$}}

\def\kzpiz{\mbox{$K^0 \pi^0$}}
\def\pippim{\mbox{$\pi^+ \pi^-$}}
\def\pippiz{\mbox{$\pi^+ \pi^0$}}

\def\pizpiz{\mbox{$\pi^0 \pi^0$}}
\def\kpkm{\mbox{$K^+ K^-$}}
\def\kpkzb{\mbox{$K^+ \overline{K}{}^0$}}

\def\kskp{\mbox{$\ks K^+$}}

\def\kzkzb{\mbox{$K^0 \overline{K}{}^0$}}
\def\ksks{\mbox{$\ks\ks$}}

\def\dzbpip{\mbox{$\dzb\pi^+$}}

\begin{document}

\vspace*{-1.5cm}
\resizebox{!}{3cm}{\includegraphics{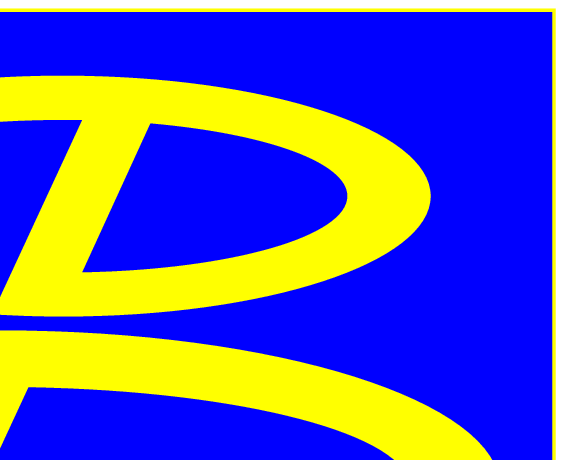}}
\preprint{\tighten\vbox{\hbox{\hfil KEK Preprint 2003-87}}}
\preprint{\tighten\vbox{\hbox{\hfil Belle Preprint 2003-26}}}

\title{\quad\\[0.5cm] \boldmath
Improved Measurements of Branching Fractions for\\
$B\to K\pi$, $\pi\pi$ and $\kk$ Decays
}

\tighten

\begin{abstract}
We report improved measurements of branching fractions for
$B\to K\pi$, $\pippim$, $\pippiz$ and $\kk$ decays based on
a data sample of $\nbb$ million $\bb$ pairs collected at the
$\Upsilon (4S)$ resonance with the Belle detector at the KEKB
$e^+e^-$ storage ring. This data sample is almost three times
larger than the sample previously used. We observe clear signals
for $B\to K\pi$, $\pippim$ and $\pippiz$ decays and set upper
limits on $B\to\kk$ decays. The results can be used to give
model-dependent constraints on the CKM angle $\phi_3$, as well as
limits on the hadronic uncertainty in the time-dependent analysis
of the angle $\phi_2$.
\end{abstract}


\affiliation{Budker Institute of Nuclear Physics, Novosibirsk}
\affiliation{Chiba University, Chiba}
\affiliation{University of Cincinnati, Cincinnati, Ohio 45221}
\affiliation{University of Frankfurt, Frankfurt}
\affiliation{Gyeongsang National University, Chinju}
\affiliation{University of Hawaii, Honolulu, Hawaii 96822}
\affiliation{High Energy Accelerator Research Organization (KEK), Tsukuba}
\affiliation{Hiroshima Institute of Technology, Hiroshima}
\affiliation{Institute of High Energy Physics, Chinese Academy of Sciences, Beijing}
\affiliation{Institute of High Energy Physics, Vienna}
\affiliation{Institute for Theoretical and Experimental Physics, Moscow}
\affiliation{J. Stefan Institute, Ljubljana}
\affiliation{Kanagawa University, Yokohama}
\affiliation{Korea University, Seoul}
\affiliation{Kyungpook National University, Taegu}
\affiliation{Swiss Federal Institute of Technology of Lausanne, EPFL, Lausanne}
\affiliation{University of Ljubljana, Ljubljana}
\affiliation{University of Maribor, Maribor}
\affiliation{University of Melbourne, Victoria}
\affiliation{Nagoya University, Nagoya}
\affiliation{Nara Women's University, Nara}
\affiliation{National Lien-Ho Institute of Technology, Miao Li}
\affiliation{Department of Physics, National Taiwan University, Taipei}
\affiliation{H. Niewodniczanski Institute of Nuclear Physics, Krakow}
\affiliation{Nihon Dental College, Niigata}
\affiliation{Niigata University, Niigata}
\affiliation{Osaka City University, Osaka}
\affiliation{Osaka University, Osaka}
\affiliation{Panjab University, Chandigarh}
\affiliation{Peking University, Beijing}
\affiliation{Princeton University, Princeton, New Jersey 08545}
\affiliation{RIKEN BNL Research Center, Upton, New York 11973}
\affiliation{University of Science and Technology of China, Hefei}
\affiliation{Seoul National University, Seoul}
\affiliation{Sungkyunkwan University, Suwon}
\affiliation{University of Sydney, Sydney NSW}
\affiliation{Tata Institute of Fundamental Research, Bombay}
\affiliation{Toho University, Funabashi}
\affiliation{Tohoku Gakuin University, Tagajo}
\affiliation{Tohoku University, Sendai}
\affiliation{Department of Physics, University of Tokyo, Tokyo}
\affiliation{Tokyo Institute of Technology, Tokyo}
\affiliation{Tokyo Metropolitan University, Tokyo}
\affiliation{Tokyo University of Agriculture and Technology, Tokyo}
\affiliation{Toyama National College of Maritime Technology, Toyama}
\affiliation{University of Tsukuba, Tsukuba}
\affiliation{Virginia Polytechnic Institute and State University, Blacksburg, Virginia 24061}
\affiliation{Yokkaichi University, Yokkaichi}
\affiliation{Yonsei University, Seoul}
  \author{Y.~Chao}\affiliation{Department of Physics, National Taiwan University, Taipei} 
  \author{K.~Suzuki}\affiliation{High Energy Accelerator Research Organization (KEK), Tsukuba} 
  \author{Y.~Unno}\affiliation{Chiba University, Chiba} 
  \author{K.~Abe}\affiliation{High Energy Accelerator Research Organization (KEK), Tsukuba} 
  \author{K.~Abe}\affiliation{Tohoku Gakuin University, Tagajo} 
  \author{T.~Abe}\affiliation{High Energy Accelerator Research Organization (KEK), Tsukuba} 
  \author{I.~Adachi}\affiliation{High Energy Accelerator Research Organization (KEK), Tsukuba} 
  \author{H.~Aihara}\affiliation{Department of Physics, University of Tokyo, Tokyo} 
  \author{M.~Akatsu}\affiliation{Nagoya University, Nagoya} 
  \author{Y.~Asano}\affiliation{University of Tsukuba, Tsukuba} 
% ...
\collaboration{The Belle Collaboration}

\maketitle

\tighten

{\renewcommand{\thefootnote}{\fnsymbol{footnote}}}
\setcounter{footnote}{0}

Recent studies at $B$ factories have significantly improved our
knowledge of heavy-flavor physics. In particular, the establishment
of mixing-induced $CP$ violation in the $B$-meson
system~\cite{phi1_belle,phi1_babar} is encouraging for further tests of
the Standard Model based on determinations of the Cabibbo-Kobayashi-Maskawa
(CKM) matrix elements~\cite{ckm}.

$B$-meson decays to $K\pi$, $\pi\pi$ and $\kk$ final states are dominated
by $b\to u$ tree and $b\to s$, $d$ penguin diagrams. The properties of
these decays provide information that can be used to determine the CKM
angles $\phi_2$ and $\phi_3$~\cite{pdg_review}.
However, the extraction of these angles suffers from hadronic
uncertainties present in the current theoretical description and from
the small amplitudes of $b\to u$, $s$, $d$ transitions. To solve these
difficulties, various theoretical approaches based on flavor symmetries
and dynamical calculations in the heavy-quark limit~\cite{hh_theory}
have been proposed. In order to utilize these methods, the precision
of the existing experimental
results~\cite{hh_belle,kspi_belle,phi2_belle_78,phi2_belle_140,hh_babar,phi2_babar,hh_cleo} must be improved.

In this paper, we report updated measurements of the branching
fractions for $B\to K\pi$, $\pippim$, $\pippiz$ and $\kk$ decays.
Recent results for $\bz\to\pizpiz$ have been reported
elsewhere~\cite{pizpiz_belle,pizpiz_babar}. The measurements reported
here are based on a 78 $\fb$ data sample collected at the $\Upsilon (4S)$
resonance, with the Belle detector~\cite{belle} at the KEKB $e^+e^-$
storage ring~\cite{kekb}. This sample
corresponds to $\Nbb$ million $\bb$ pairs and is about three
times larger than that used for our previous analysis~\cite{hh_belle}.
The previous results are superseded with significantly improved
statistical precision. Throughout this paper, neutral and charged
$B$ mesons are assumed to be produced in equal amounts at the
$\Upsilon (4S)$. The inclusion of the charge conjugate decay is
implied, unless explicitly stated.

The Belle detector is a large-solid-angle spectrometer consisting
of a three-layer silicon vertex detector, a 50-layer central drift
chamber (CDC), an array of threshold Cherenkov counters with silica
aerogel radiators (ACC), time-of-flight scintillation counters, and
an electromagnetic calorimeter comprised of CsI(Tl) crystals (ECL)
located inside a superconducting solenoid coil that provides a 1.5 T
magnetic field. An iron flux-return located outside of the coil is
instrumented to detect $K^0_L$ mesons and to identify muons. A detailed
description of the Belle detector can be found elsewhere~\cite{belle}.

The basic analysis procedure is the same as described in
Ref.~\cite{hh_belle}.
However, the data sample used in this analysis was reprocessed with an
improved tracking algorithm that reduces the probability of incorrectly
associating CDC hits in the track finding. This improvement changes the
efficiencies for the kinematic reconstruction of the signal as well as
for the measurement of specific ionization energy loss ($dE/dx$) in the
CDC from the values given in Ref.~\cite{hh_belle}.

The $\pi^{\pm}$ mass is assigned to each charged
track. Tracks used to form $B$ candidates are required to originate from
the interaction region based on their impact parameters. $\ks$ mesons
are reconstructed using pairs of oppositely charged tracks that have
invariant masses in the range $480$~MeV/$c^2 <M_{\pi\pi}< 516$~MeV/$c^2$.
A reconstructed $\ks$ is required to have a displaced vertex and a flight
direction consistent with that of a $\ks$ originating from the interaction
region. Pairs of photons with invariant masses in
the range $115$~MeV/$c^2 < M_{\gamma\gamma} < 152$~MeV/$c^2$ are used
to form $\pi^0$ mesons. The measured energy of each photon in the
laboratory frame is required to be greater than $50$~MeV in the barrel
region, defined as $32^{\circ} < \theta_{\gamma} < 128^{\circ}$, and
greater than $100$~MeV in the end-cap regions, defined as
$17^{\circ}\le\theta_{\gamma}\le 32^{\circ}$ or
$128^{\circ}\le\theta_{\gamma}\le 150^{\circ}$, where
$\theta_{\gamma}$ denotes the polar angle of the photon with respect to
the $e^-$ beam.
Signal $B$ candidates are required to satisfy
$5.27$~GeV/$c^2 < \mbc < 5.29$~GeV/$c^2$ and $-0.3$~GeV$ < \de < 0.5$~GeV,
where $\mbc=\sqrt{E_{\rm beam}^{*2}-p_B^{*2}}$,
$\de = E_B^* - E_{\rm beam}^*$, $E_{\rm beam}^*$ is the beam-energy, and
$p_B^*$ and $E_B^*$ are the momentum and energy of the reconstructed $B$
meson, all evaluated in the $e^+e^-$ center-of-mass (CM) frame.
The signal efficiencies of the kinematic reconstruction, estimated using
GEANT-based~\cite{geant} Monte Carlo (MC) simulations, are listed
in Table~\ref{tab:eff}.

Charged tracks from $B$ candidates have momenta ranging from $1.5$
up to $4.5$~GeV/$c$ in the laboratory frame. They are distinguished
as $K^{\pm}$ or $\pi^{\pm}$ mesons by the number of photoelectrons
($N_{\rm p.e.}$) detected by the ACC and
$dE/dx$ measured in the CDC.
These quantities are used to form
a $K^{\pm}$ identification (KID) likelihood ratio
${\cal R}_K = {\cal L}_K / ( {\cal L}_K + {\cal L}_{\pi} )$,
where ${\cal L}_K$ denotes the product of the individual likelihoods
of $N_{\rm p.e.}$ and $dE/dx$ for $K^{\pm}$ mesons, and ${\cal L}_{\pi}$
is the corresponding product for $\pi^{\pm}$ mesons.
The requirements on ${\cal R}_K$ used in this analysis yield a $K^{\pm}$
identification efficiency of 84.4\% with a $\pi^{\pm}$ misidentification
rate of 5.3\% for $K^{\pm}$ candidates, and a $\pi^{\pm}$ identification
efficiency of 91.2\% with a $K^{\pm}$ misidentification rate of 10.2\% for
$\pi^{\pm}$ candidates.
The efficiencies and misidentification rates are measured by comparing
the yields of high-momentum $D^{*+}$-tagged $\dz\to\kmpip$ decays before
and after applying the ${\cal R}_K$ requirements.
Here, the $K^{\pm}$ and $\pi^{\pm}$ momentum range is required to be the
same as for the signal. Since the momentum and angular distributions are
slightly different for $\dz$ data and signal MC, the KID efficiencies are
reweighted as a function of the polar angle of the signal track with
respect to the $e^-$ beam.
In addition to the KID requirement, positively identified electrons are
rejected using a similar likelihood ratio that also includes the energy
deposited in the ECL.

The dominant background is due to the $e^+e^-\to\qq$ ($q = u$, $d$, $s$,
$c$) continuum processes. A large MC sample shows that backgrounds from
the $b\to c$ transition are negligible since the momenta of their decay
products are smaller than those in the signal decays. On the other hand,
the momenta of the decay products from $b\to u$, $s$, $d$ transitions
other than the signal (denoted as other charmless $B$ decays) can be as
large as those in the signal decays. Events from these charmless $B$
decays populate the negative $\de$ region because of the energy carried
away by a photon or $\pi$ meson, which is not used in the $B$ reconstruction.
We take these events into account in the signal extraction as discussed later.

We discriminate signal events from the $\qq$ background by the event
topology. This is quantified by the Super-Fox-Wolfram ($SFW$)
variable~\cite{hh_belle}, which is a Fisher discriminant~\cite{fisher}
formed from modified Fox-Wolfram moments~\cite{fw}.
The angle of the $B$-meson flight direction with respect to the beam
axis in the CM frame ($\theta_B$) provides additional discrimination.
A signal likelihood ratio ${\cal R}_s = {\cal L}_s / ({\cal L}_s +
{\cal L}_{q\overline{q}})$ is used as the discriminating variable,
where ${\cal L}_s$ denotes the product of the individual $SFW$ and
$\theta_B$ likelihoods for the signal, and ${\cal L}_{q\overline{q}}$
is that for the $\qq$ background.
The probability density functions (PDFs) used for the likelihoods are
derived from the MC for the signal, while events in the $\mbc$ sideband
($5.2$~GeV/$c^2 < \mbc < 5.26$~GeV/$c^2$ in the $\de$ acceptance) are
used for the $\qq$ background.
We make a mode-dependent requirement on ${\cal R}_s$ that maximizes
$N_s^{\rm exp} / \sqrt{N_s^{\rm exp} + N_{q\overline{q}}^{\rm exp}}$,
where $N_s^{\rm exp}$ and $N_{q\overline{q}}^{\rm exp}$ denote the
expected signal and $\qq$ yields based on our previous
measurements~\cite{hh_belle} (upper limits are used for $\kk$ modes).
The ${\cal R}_s$ requirements eliminate more than 90\% of the $\qq$
background for the signal efficiencies given in Table~\ref{tab:eff}.

\begin{figure*}[!htb]
\centerline{
\epsfig{file=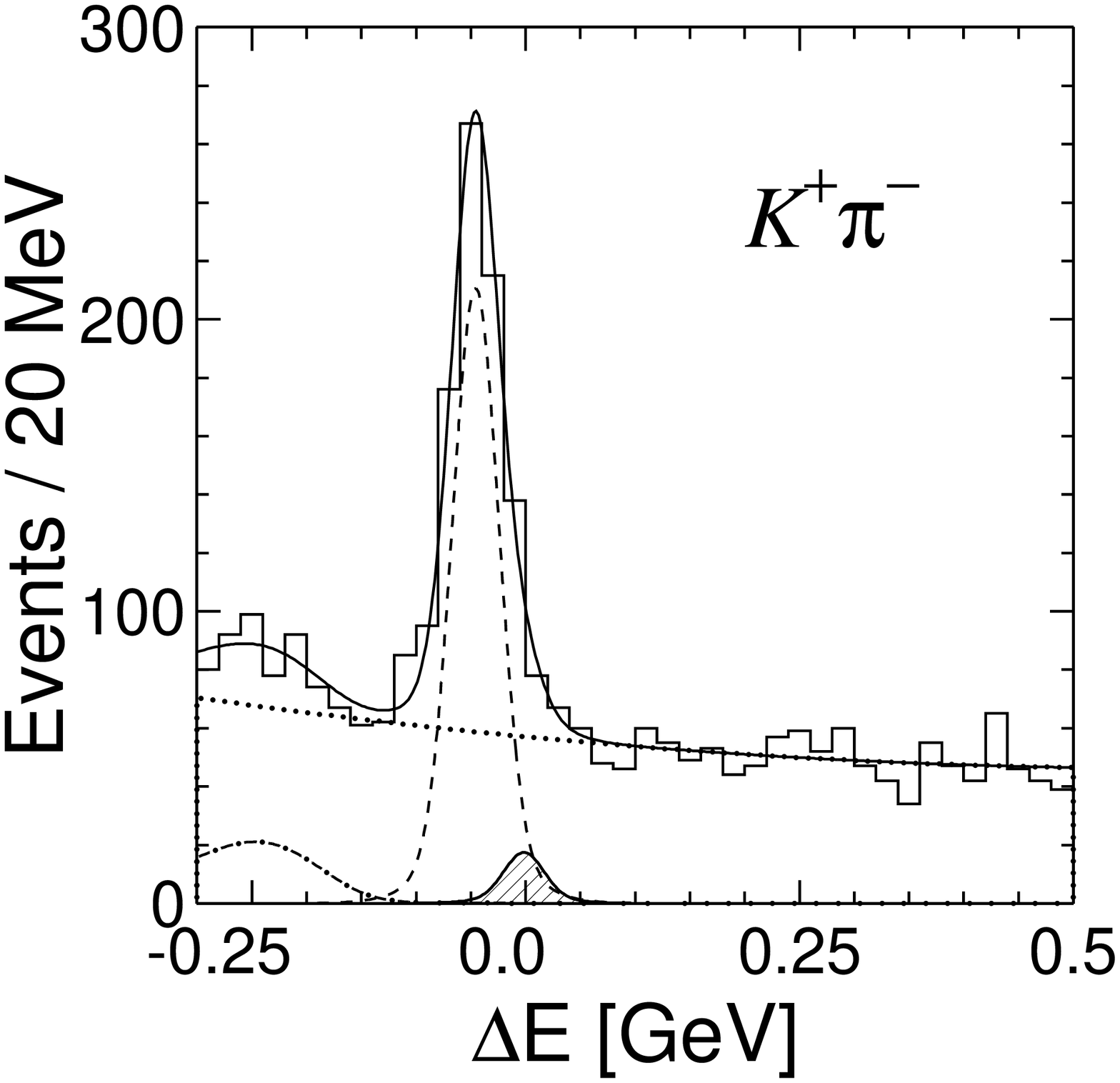,width=3.7cm}
\hspace{-4mm}
\epsfig{file=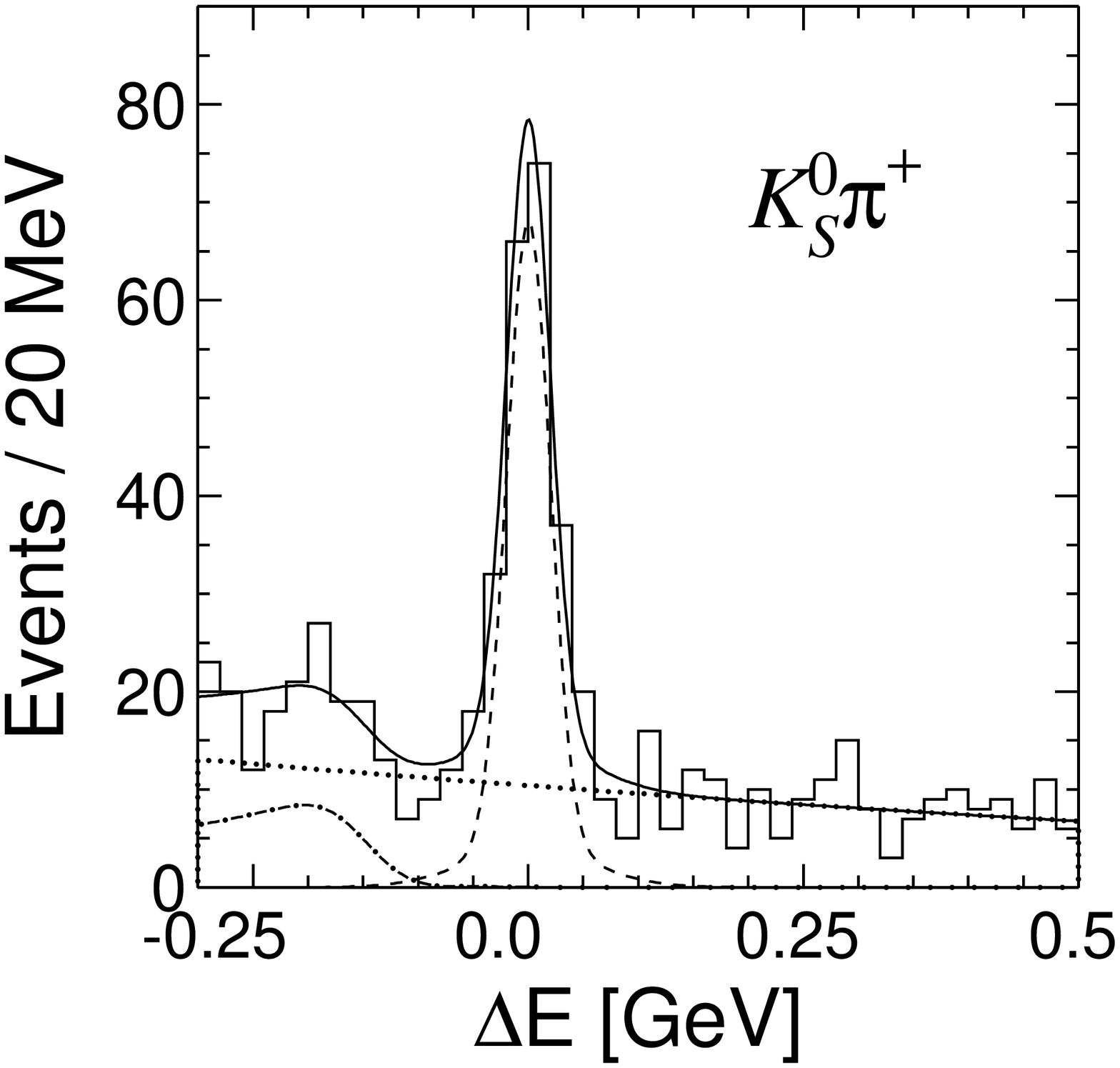,width=3.7cm}
\hspace{-4mm}
\epsfig{file=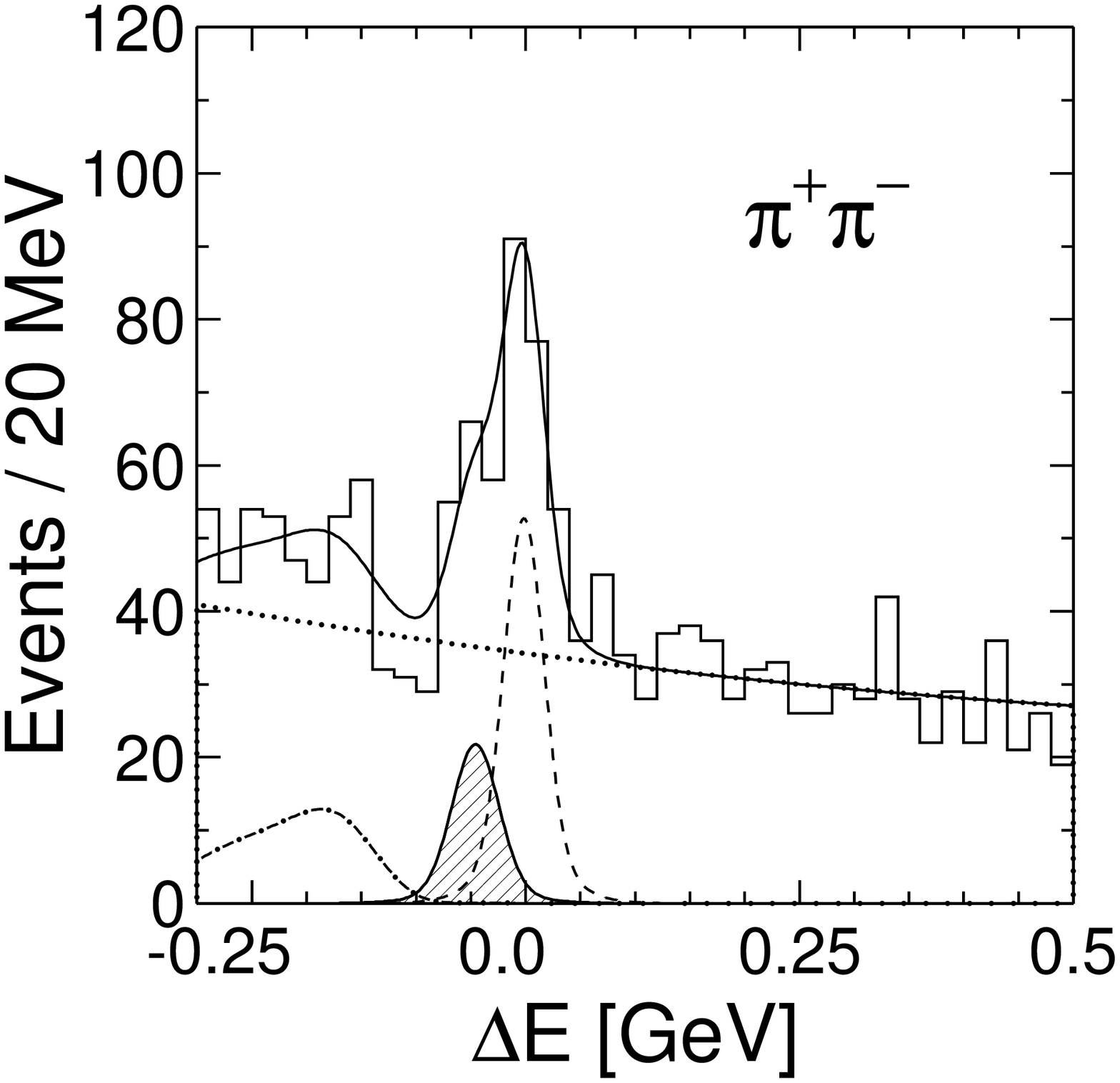,width=3.7cm}
\hspace{-4mm}
\epsfig{file=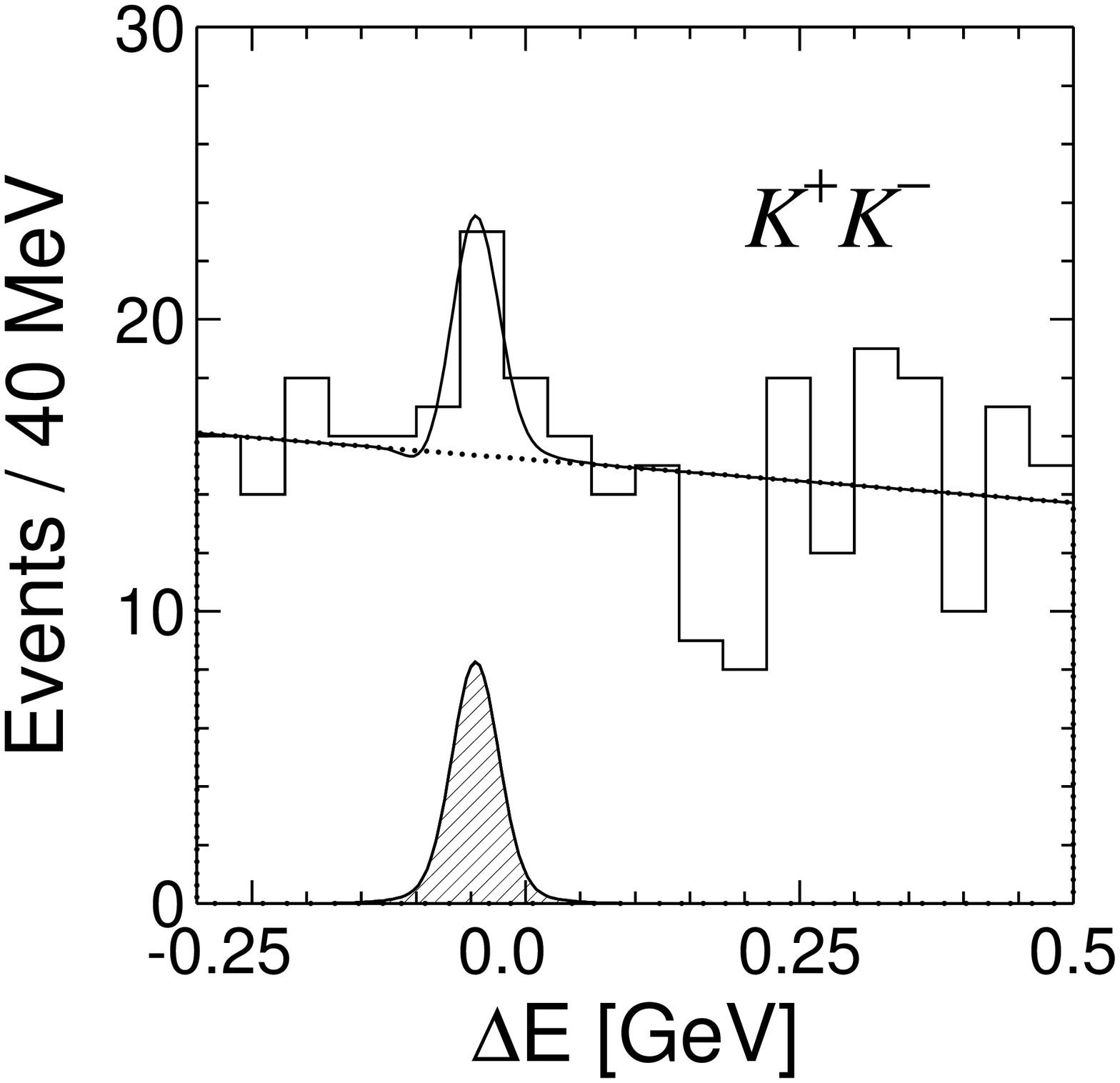,width=3.7cm}
\hspace{-4mm}
\epsfig{file=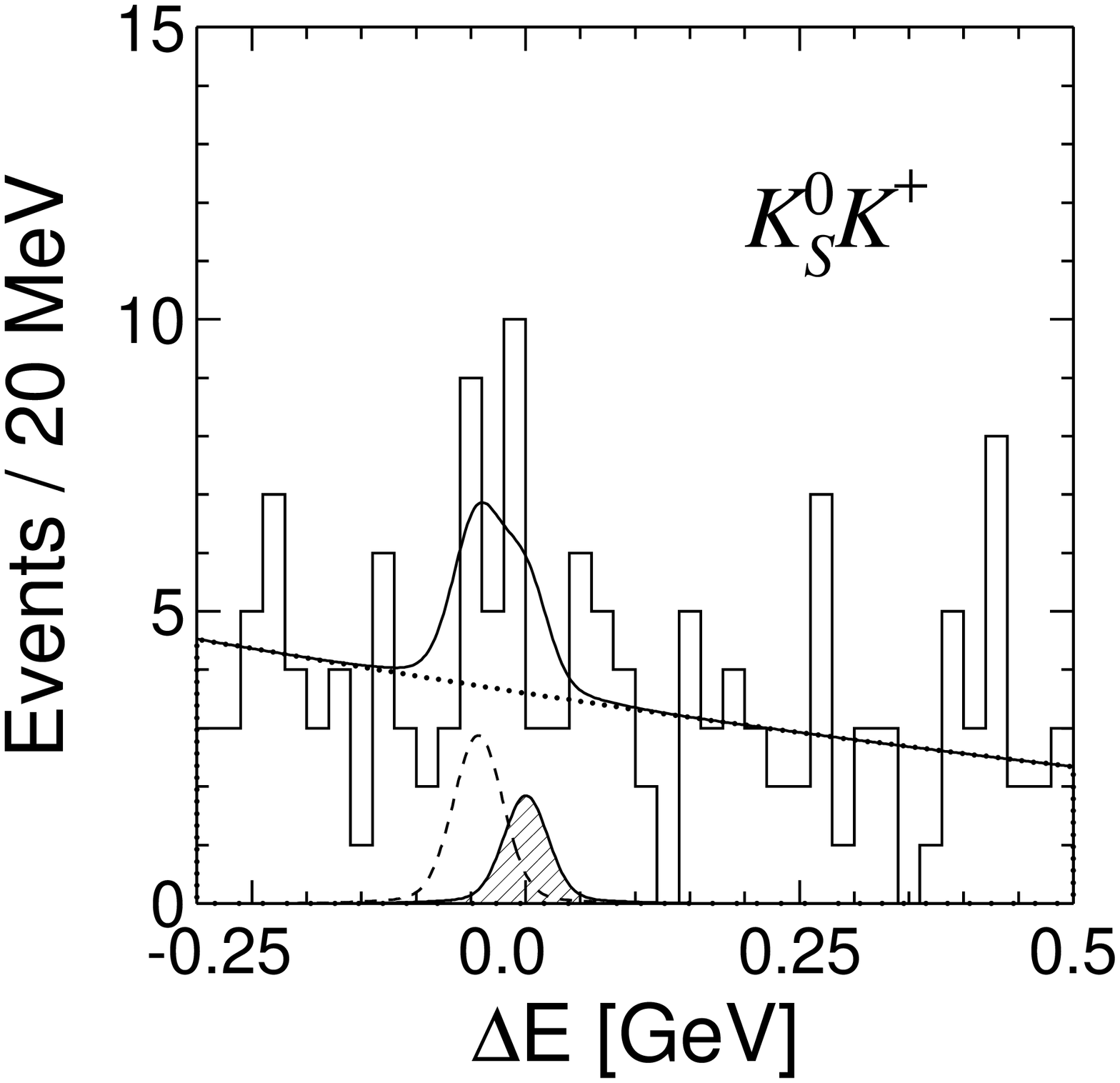,width=3.7cm}
}
\centerline{
\epsfig{file=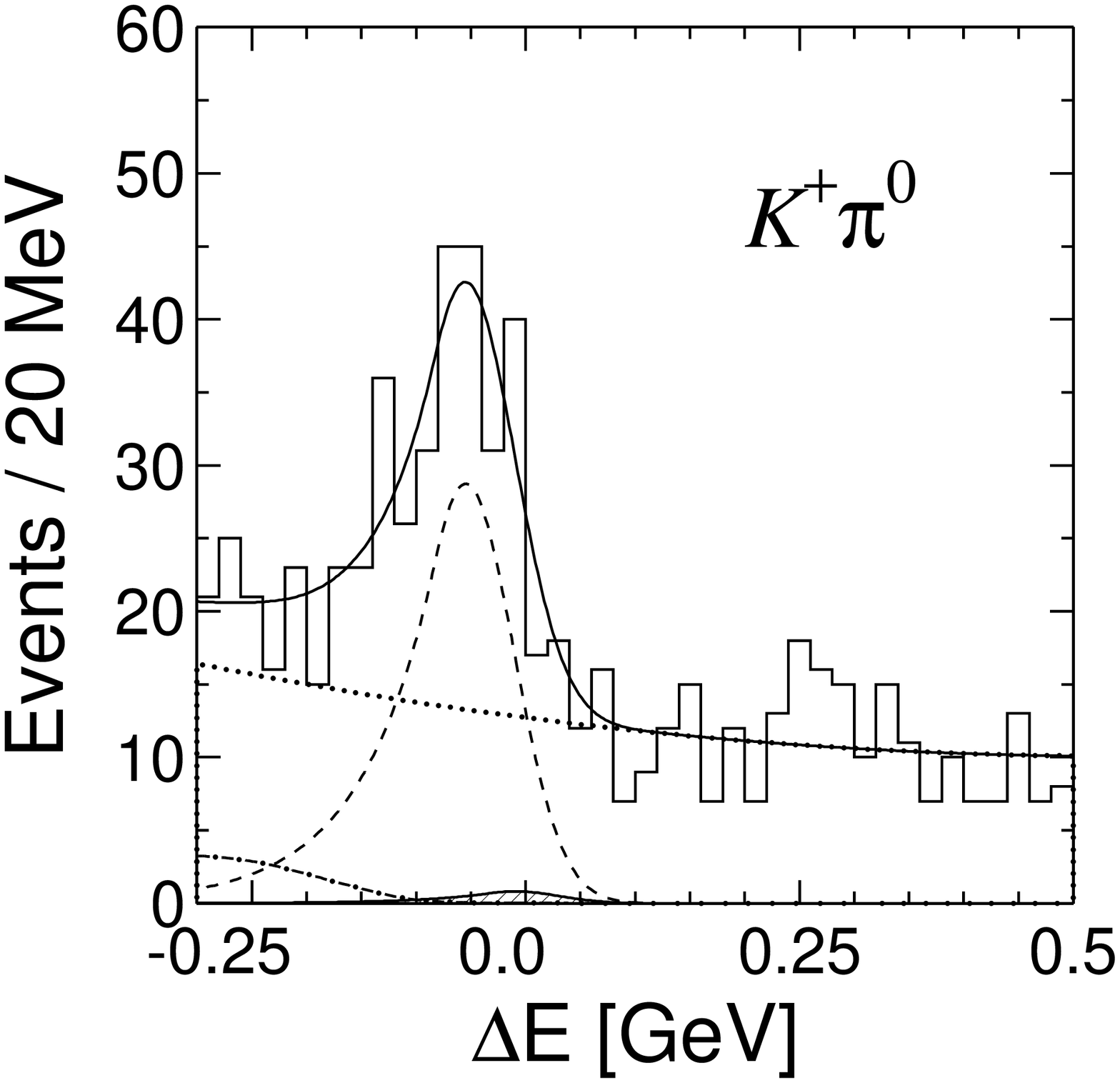,width=3.7cm}
\hspace{-4mm}
\epsfig{file=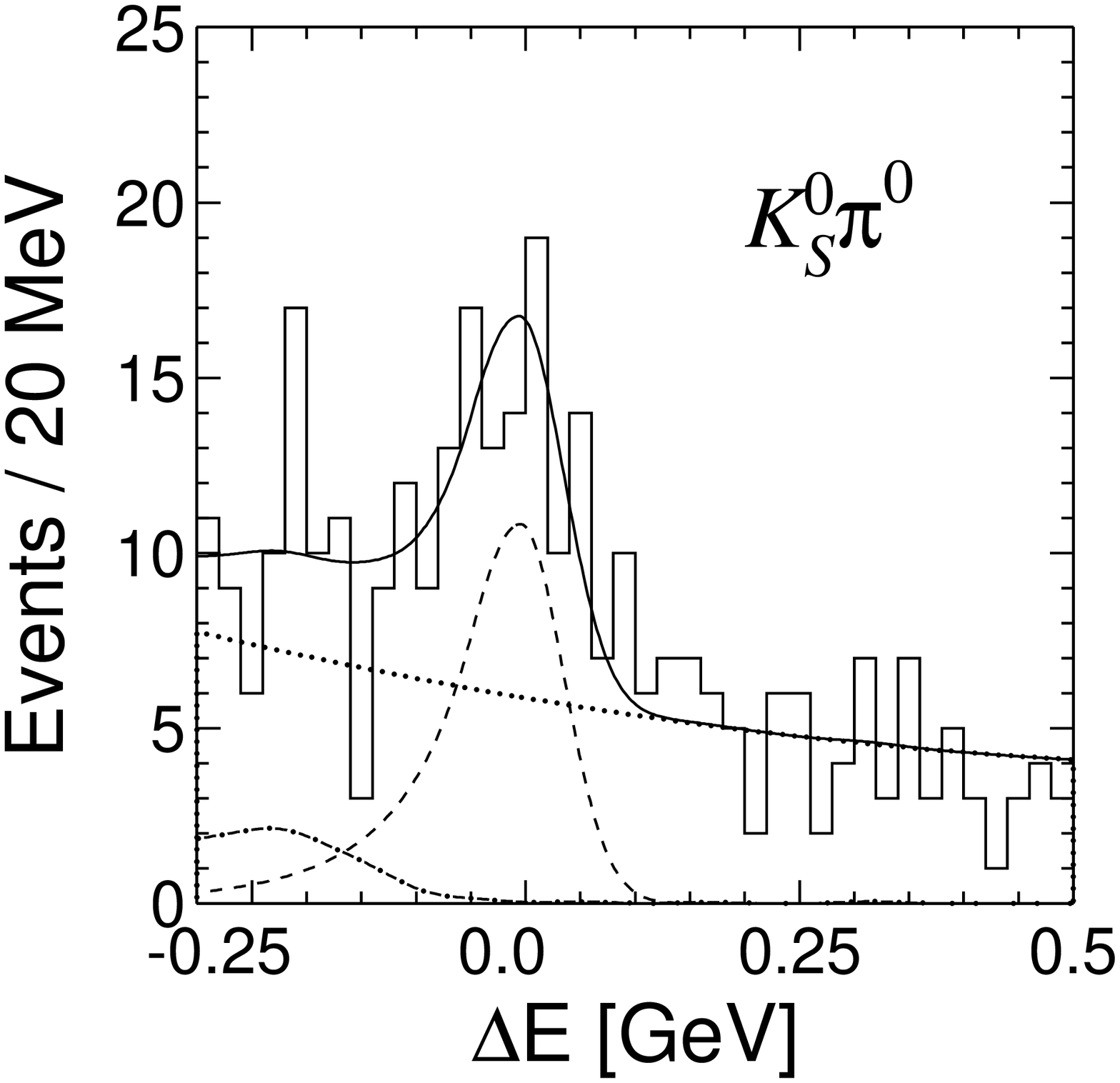,width=3.7cm}
\hspace{-4mm}
\epsfig{file=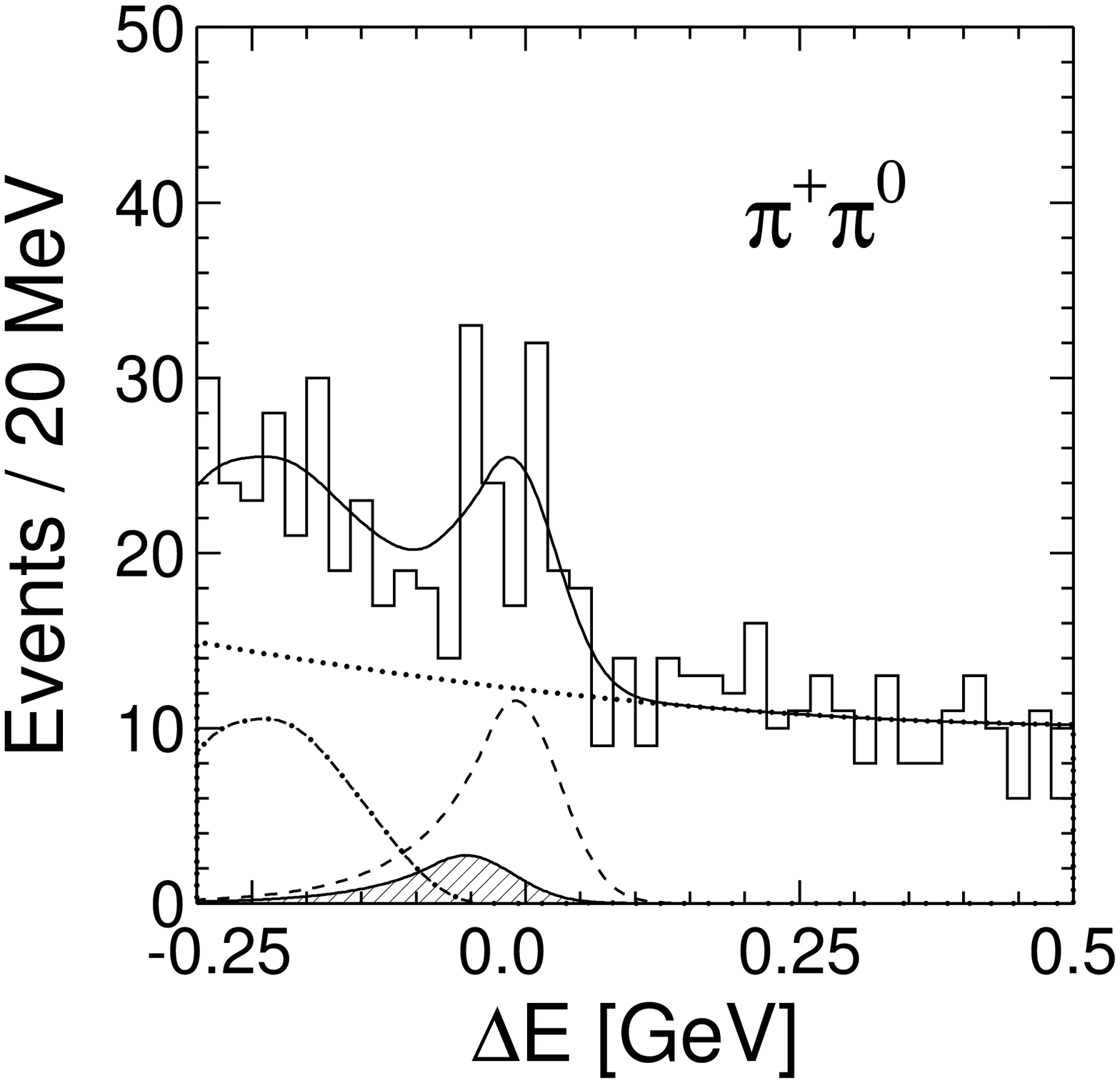,width=3.7cm}
\hspace{-4mm}
\epsfig{file=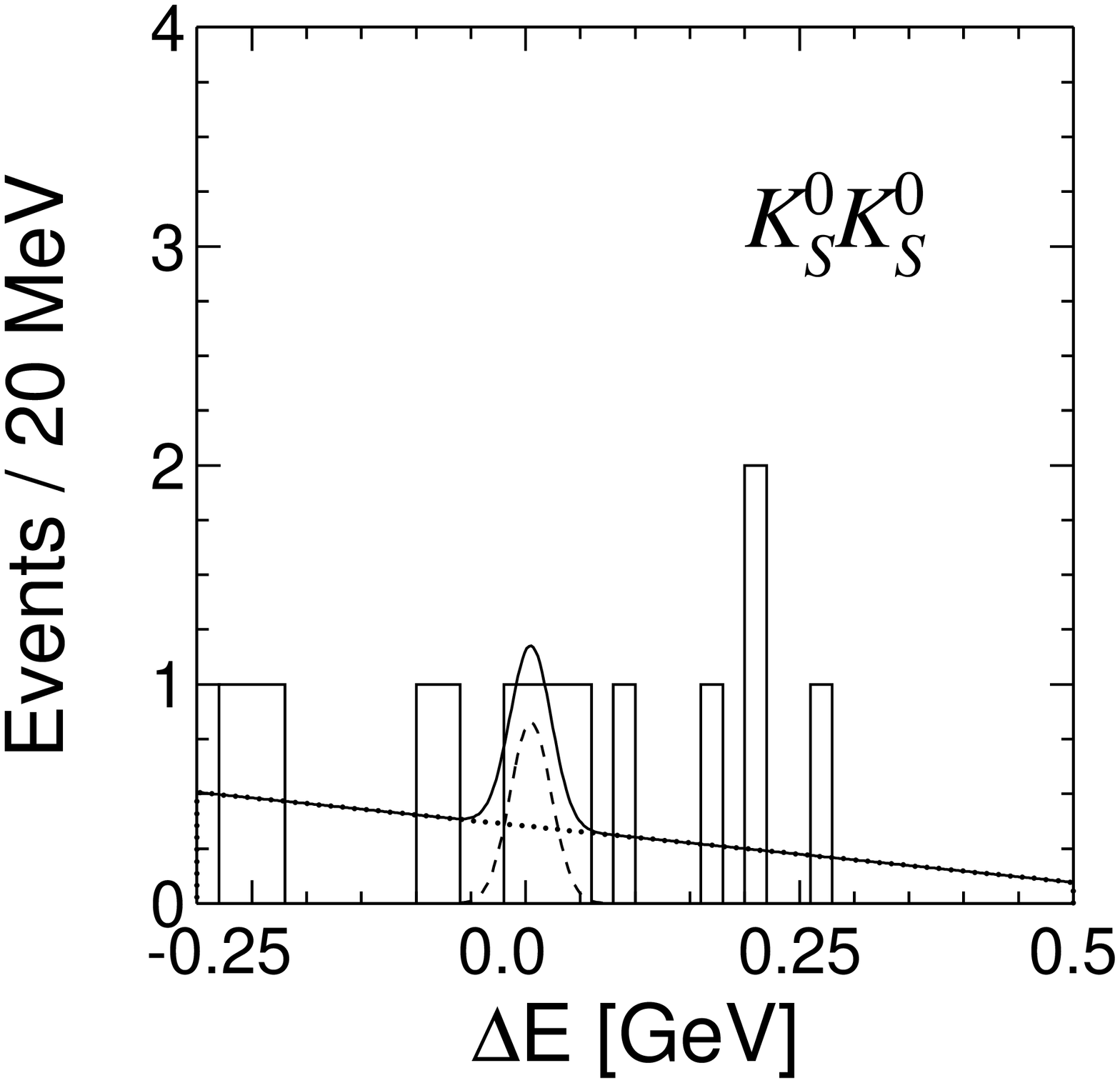,width=3.7cm}
\hspace{-4mm}
\hspace{3.7cm}
}
\caption{
$\de$ distributions for $B\to K\pi$, $\pippim$, $\pippiz$ and $\kk$ decays.
Fit results are shown as the solid, dashed, dotted and dot-dashed curves
for the total, signal, $\qq$ background and the other charmless $B$ decays,
respectively. In addition, reflections due to $K^{\pm}/\pi^{\pm}$
misidentification are shown as hatched areas.
}
\label{fig:br_hh}
\end{figure*}

Signal yields are extracted using a binned maximum-likelihood fit to
the $\de$ distributions after all the event selection requirements
discussed above. The fitting function contains components for the
signal, $\qq$ background, and other charmless $B$ decays.
If applicable, possible reflections due to the $K^{\pm}/\pi^{\pm}$
misidentification are included as additional components.
All of the fit parameters other than the normalizations are fixed.
The signal PDFs are based on the MC. For the modes with a $\pi^0$ meson,
the PDF is modeled with an empirically determined
parametrization~\cite{cbline}. For the other modes, the sum of two
Gaussian distributions with a common mean is used for the PDF.
Due to the $\pi^{\pm}$ mass assumption, each $K^{\pm}$ meson in the
final state results in a shift in the peak position of about $-45$~MeV.
Discrepancies between the peak positions in data and MC are calibrated
using $\bp\to\dzbpip$ decays, where the $\dzb\to\kppim\pi^0$ sub-decay
is used for the modes with $\pi^0$ mesons and the $\dzb\to\kppim$
sub-decay is used for the other modes. Here, the same analysis
procedure used for the signal is applied except for the daughter
particle reconstruction. The MC-based $\de$ resolutions are calibrated
using invariant mass resolutions of high-momentum inclusive $D$ decays.
We use $\dz\to\kmpip$ for the $\bz\to\kppim$, $\pippim$, and $\kpkm$ modes,
$\dplus\to\kspip$ for the $\bp\to\kspip$, $\kskp$ and $\bz\to\ksks$ modes,
and $\dz\to\kmpip\pi^0$ for the modes with a $\pi^0$ meson.
The momentum ranges and reconstruction procedures for the $D$ daughter
particles are required to be the same as those for the signal daughter
particles. The signal PDFs are also used for the reflections. The PDF
of the $\qq$ background is determined from the $\mbc$ sideband data and
modeled with a first-order polynomial for the $\kpkm$ and $\ksks$ modes
and a second-order polynomial for the other modes. The PDF for the other
charmless $B$ decays is taken from a smoothed histogram of a large MC
sample. The $\kppiz$ and $\pippiz$ modes are fitted simultaneously with
a fixed reflection-to-signal ratio that is determined from the measured
KID efficiencies and misidentification rates. For other modes, all the
normalizations are floated.
All fit results are shown in Fig.~\ref{fig:br_hh}.

The obtained signal yields are listed in Table~\ref{tab:br} together
with their statistical significances ${\cal S}=\sqrt{-2\ln({\cal L}_0/
{\cal L}_{N_s})}$, where ${\cal L}_0$ and ${\cal L}_{N_s}$ denote the
maximum likelihoods of the fits without and with the signal component,
respectively.
The fitted reflection yields are consistent within statistics with the
expectations, which are derived from the fitted signal yields, KID
efficiencies and misidentification rates, and the efficiencies of the
$R_s$ requirements.
We observe clear signals for $B\to K\pi$, $\pippim$ and $\pippiz$ decays.
For the decays $B\to\kk$, no significant signal is observed.
We apply the Feldman-Cousins frequentist approach with systematic
uncertainties taken into account~\cite{fc} to obtain upper limits on the
yields at the 90\% confidence level (CL); these are used to set branching
fraction upper limits.
The branching fractions and upper limits are listed in Table~\ref{tab:br}.
Here, our latest measurement for $\bz\to\pizpiz$~\cite{pizpiz_belle} based
on a data sample of 152 million $\bb$ pairs is also listed for completeness.
The hierarchy of the branching fractions,
${\cal B}(B\to K\pi)>{\cal B}(B\to\pi\pi)$, is confirmed. More statistics
are needed in order to firmly establish the position of $B\to\kk$ in this
hierarchy.

The systematic errors in the branching fractions are the quadratic sums
of the systematic errors in the signal yields, uncertainties in the
reconstruction efficiencies, and the 0.6\% error in the number of $\bb$
pairs.
The systematic errors in the signal yields come from the uncertainties
in the fit procedure.
In order to study the sensitivity to the signal and $\qq$ background
PDFs, each shape parameter is independently varied by its error in
the fit.
The sensitivity to the contribution from other charmless $B$ decays is
evaluated by changing the minimum $\de$ requirement to $-100$~MeV
($-150$~MeV) for the modes without (with) a $\pi^0$ meson, to exclude
most of these events from the fit.
The resulting changes in the signal yield are added in quadrature
and assigned to the systematic errors on the signal yields as listed
in Table~\ref{tab:br}.
The uncertainties in the reconstruction efficiencies are listed in
Table~\ref{tab:br_sys} along with the test samples that are used.
The uncertainty for the track finding efficiency in the high-momentum
region is obtained by comparing the ratio of yields of fully reconstructed
and partially reconstructed test samples in data and MC. The uncertainties
in the $\ks$ and $\pi^0$ reconstruction efficiencies are obtained
from similar comparisons of yield ratios in test samples. Here, the test
samples are restricted to the same $\ks$ and $\pi^0$ momentum ranges as
the signal.
The experimental errors in the branching fractions of these decays~\cite{pdg}
are added in quadrature.
The uncertainties in the KID efficiencies and misidentification rates
are due to the statistics of the data test sample.
We also checked the effect of the difference in the hadronic environment
between the signal and test samples.
No significant effect is seen in the efficiencies and misidentification rates.
The ${\cal R}_s$ requirement for each mode is applied to
data and MC test samples, and the difference is included in the systematic
error.

To a good approximation in the Standard Model, the relative weak phase
between the penguin and tree amplitudes in $K\pi$ modes is $\phi_3$.
It is in principle possible to extract $\phi_3$ if the hadronic
uncertainties are under control. Several approaches to constrain $\phi_3$
have been proposed using the ratios of partial widths for $K\pi$ and $\pi\pi$
modes with model-dependent assumptions on the hadronic
uncertainties~\cite{hh_theory}; the ratios give cancellations of these
uncertainties. We calculate such useful partial width ratios as listed
in Table~\ref{tab:rbr}. Here, the ratio of charged to neutral $B$ meson
lifetimes $\tau_{B^+}/\tau_{B^0}=1.083\pm 0.017$~\cite{pdg} is used to
convert the branching fraction ratios into partial width ratios if
necessary, and the total errors are reduced because of the cancellation
of the partially common systematic errors.
Applying the approach of Buras and Fleischer~\cite{bfbound}, for
illustration, our
$\Gamma(\kppim)/2\Gamma(\kzpiz)$ measurement excludes the region
$29^{\circ} < \phi_3 < 83^{\circ}$ at the 90\% CL based on MC
pseudo-experiments while that
of $2\Gamma(\kppiz)/\Gamma(\kzpip)$ gives no constraint. These results are
obtained without any assumption on the tree-to-penguin amplitude
ratio, but neglecting re-scattering effects and taking the size of the
electroweak penguin as in Ref.~\cite{bfbound}.
Although a more aggressive constraint on $\phi_3$ can be derived by
introducing further model-dependent assumptions on the hadronic
uncertainties, a coherent study of these approaches is required to
reduce the model-dependence on hadronic uncertainties and to
determine $\phi_3$.

A naive expectation for the tree-dominated $\pippim$ and $\pippiz$ modes
predicts $2\Gamma(\pippiz)/\Gamma(\pippim) = 1$. The deviation of our
result from this expectation, as given in Table~\ref{tab:rbr}, is consistent
with our previous measurement~\cite{hh_belle} and would indicate the
existence of a significant penguin contribution in the $\pippim$ mode
if the color-suppressed tree contribution plays a minor role~\cite{grreview}.
This penguin contribution complicates the extraction of $\phi_2$ from
the time-dependent $CP$ asymmetry in the $\pippim$ mode (referred to
as ``penguin pollution'')~\cite{isospin}.
Applying an approach based on the isospin relations in $\pi\pi$
modes~\cite{glbound},
our measured ratios $\Gamma(\pippiz)/\Gamma(\pippim)$ and
$\Gamma(\pizpiz)/\Gamma(\pippim)$ in Table~\ref{tab:rbr} give the 90\% CL
bound on the size of the ``penguin pollution'' $|\theta|< 56^{\circ}$; the
CL is derived from MC pseudo-experiments.
Here, we also use the partial-rate $CP$ asymmetry in the $\pippim$
mode, ${\cal A}_{\pi\pi} = +0.58\pm 0.15\pm 0.07$, which is reported in
Ref.~\cite{phi2_belle_140}.

In conclusion, we have measured or constrained the branching fractions
for the $B\to K\pi$, $\pippim$, $\pippiz$ and $\kk$ decays with $\nbb$
million $\bb$ pairs collected on the $\Upsilon (4S)$ resonance at the
Belle experiment.
We observe clear signals for $B\to K\pi$, $\pippim$ and $\pippiz$ decays
and set upper limits on $B\to\kk$ decays.
The hierarchy of branching fractions reported in earlier measurements
is confirmed.
These results have significantly improved statistical precision
compared to
our previous measurements and supersede them.
The results can be used to give model-dependent constraints on $\phi_3$,
as well as limits on the hadronic uncertainty in the time-dependent
analysis of $\phi_2$.

We wish to thank the KEKB accelerator group for the excellent
operation of the KEKB accelerator.
We acknowledge support from the Ministry of Education,
Culture, Sports, Science, and Technology of Japan
and the Japan Society for the Promotion of Science;
the Australian Research Council
and the Australian Department of Education, Science and Training;
the National Science Foundation of China under contract No.~10175071;
the Department of Science and Technology of India;
the BK21 program of the Ministry of Education of Korea
and the CHEP SRC program of the Korea Science and Engineering Foundation;
the Polish State Committee for Scientific Research
under contract No.~2P03B 01324;
the Ministry of Science and Technology of the Russian Federation;
the Ministry of Education, Science and Sport of the Republic of Slovenia;
the National Science Council and the Ministry of Education of Taiwan;
and the U.S.\ Department of Energy.

\begin{table}[!htb]
\caption{Signal efficiencies for kinematic reconstruction (Rec),
${\cal R}_K$ requirements and ${\cal R}_s$ requirement along with
the sub-decay branching fraction (${\cal B}_{\rm sub}$) for
$K^0\to\ks\to\pippim$ and total signal efficiencies.
}
\label{tab:eff}
\begin{ruledtabular}
\begin{tabular}{lccccc}
\cc{Mode} & Rec & ${\cal R}_K$ & ${\cal R}_s$
	& ${\cal B}_{\rm sub}$ & Total\\
\hline			     
\kppim  & 0.731 & 0.769 & 0.672 & ----- & 0.378 \\
\kppiz 	& 0.461 & 0.844 & 0.501 & ----- & 0.195 \\
\kzpip 	& 0.571 & 0.911 & 0.560 & 0.343 & 0.100 \\
\kzpiz 	& 0.314 & ----- & 0.673 & 0.343 & 0.073 \\
\pippim & 0.756 & 0.830 & 0.560 & ----- & 0.352 \\
\pippiz & 0.476 & 0.911 & 0.395 & ----- & 0.172 \\
\kpkm   & 0.727 & 0.713 & 0.387 & ----- & 0.201 \\
\kpkzb 	& 0.539 & 0.844 & 0.388 & 0.343 & 0.061 \\
\kzkzb 	& 0.447 & ----- & 0.561 & 0.235 & 0.059 \\
\end{tabular}
\end{ruledtabular}
\end{table}
\begin{table}[!htb]
\caption{Signal yields ($N_s$), statistical significance (${\cal S}$),
and branching fractions ($\br$) for the $B\to K\pi$, $\pippim$, $\pippiz$
and $\kk$ decays. The first and second errors are the statistical and
systematic errors, respectively.
For completeness, the $\pizpiz$ results from Ref.~\cite{pizpiz_belle}
are also listed.
}
\label{tab:br}
\begin{ruledtabular}
\begin{tabular}{llll}
\cc{Mode}	& \cc{$N_s$}	& \cc{${\cal S}$ [$\sigma$]}
	& \cc{${\cal B}$ [$10^{-6}$]}\\
\hline
$\kppim$	& \NsA	& \SigA	& \BrA\\
$\kppiz$	& \NsB	& \SigB	& \BrB\\
$\kzpip$	& \NsC	& \SigC	& \BrC\\
$\kzpiz$	& \NsD	& \SigD	& \BrD\\
\hline				
$\pippim$	& \NsE	& \SigE	& \BrE\\
$\pippiz$	& \NsF	& \SigF	& \BrF\\
$\pizpiz$	& \NsG	& \SigG	& \BrG\\
\hline				
$\kpkm$		& \NsH	& \SigH	& \cc{\BrH}\\
$\kpkzb$	& \NsI	& \SigI	& \cc{\BrI}\\
$\kzkzb$	& \NsJ	& \SigJ	& \cc{\BrJ}\\
\end{tabular}
\end{ruledtabular}
\end{table}
\begin{table}[!htb]
\caption{Uncertainties on the reconstruction efficiencies
($\delta\epsilon$) along with the test samples that are used.}
\label{tab:br_sys}
\begin{ruledtabular}
\begin{tabular}{ccl}
Source	& $\delta\epsilon/\epsilon$ [\%] & \cc{Test sample} \\
\hline
track finding
	& $1.0$
	& $D^{*+}\to \dz(\to\ks\pippim)\pi^+$\\
$\ks$
	& $4.4$
	& $D^+\to\kspip$, $\kmpip\pi^+$\\
$\pi^0$
	& $3.5$
	& $\eta\to\pizpiz\pi^0$, $\gamma\gamma$\\
${\cal R}_K$
	& $0.2$
	& $D^{*+}\to\dz(\to\kmpip)\pi^+$\\
${\cal R}_s$
	& $1.3$--$7.8$
	& $\bp\to\dzb(\to\kppim$, $\kppim\pi^0)\pi^+$\\
\end{tabular}
\end{ruledtabular}
\end{table}
\begin{table}[!htb]
\caption{Partial width ratios of $B\to K\pi$ and $\pi\pi$
decays. The errors are quoted in the same manner as in
Table~\ref{tab:br}.}
\label{tab:rbr}
\begin{ruledtabular}
\begin{tabular}{ll}
\cc{Modes} & \cc{Ratio}\\
\hline
$\hdig \Gamma(\kppim)       / \hdig \Gamma(\kzpip)$
	& $0.91 \pm 0.09 \pm 0.06$\\
$\hdig \Gamma(\kppim)       / 2     \Gamma(\kzpiz)$
	& $0.79 \pm 0.16 \pm 0.09$\\
$2     \Gamma(\kppiz)  \hsb / \hdig \Gamma(\kzpip)$
	& $1.09 \pm 0.15 \aer{0.13}{0.10}$\\
$\hdig \Gamma(\pippim) \hsa / \hdig \Gamma(\kppim)$
	& $0.24 \pm 0.03 \pm 0.02$\\
$\hdig \Gamma(\pippiz) \hse / \hdig \Gamma(\kzpiz)$
	& $0.39 \pm 0.12 \pm 0.06$\\
$2     \Gamma(\pippiz) \hse / \hdig \Gamma(\kzpip)$
	& $0.45 \pm 0.12 \pm 0.05$\\
$2     \Gamma(\pippiz) \hse / \hdig \Gamma(\pippim)$
	& $2.10 \pm 0.58 \pm 0.25$\\
$\hdig \Gamma(\pizpiz) \hsg / \hdig\hsh \Gamma(\pippim)$
	& $0.39 \pm 0.15 \pm 0.05$\\
$\hdig \Gamma(\pizpiz) \hsg / \hdig\hsh \Gamma(\pippiz)$
	& $0.37 \pm 0.16 \pm 0.05$\\
\end{tabular}
\end{ruledtabular}
\end{table}

\end{document}